\documentclass[twocolumn]{aastex62} 

\usepackage{xspace}
\usepackage{graphicx}
\usepackage{natbib}
\usepackage{amsmath}
\usepackage{booktabs}
\usepackage{multirow}
\usepackage{afterpage}
\usepackage{rotating}

\newcommand{\nraoblurb}{The National Radio Astronomy Observatory is
a facility of the National Science Foundation operated under cooperative
agreement by Associated Universities, Inc.}
\newcommand{\hide}[1]{}

\newcommand{\lsim}{\ensuremath{\,\lesssim\,}\xspace}
\newcommand{\kms}{\ensuremath{\,{\rm km\,s^{-1}}}\xspace}

\newcommand{\cm}{\ensuremath{\,{\rm cm}}\xspace}
\newcommand{\pc}{\ensuremath{\,{\rm pc}}\xspace}
\newcommand{\kpc}{\ensuremath{\,{\rm kpc}}\xspace}
\newcommand{\K}{\ensuremath{\,{\rm K}}\xspace}

\newcommand{\ghz}{\ensuremath{\,{\rm GHz}}\xspace}

\newcommand{\degree}{\ensuremath{^\circ}\xspace}

\newcommand{\msun}{\ensuremath{\,M_\odot}\xspace}     
 
\newcommand{\hii}{{\rm H\,{\footnotesize II}}\xspace}

\shorttitle{Galactic Center Lobe}
\shortauthors{Anderson et al.}

\begin{document}

\title{The Galactic Center Lobe as an H{\small II} Region}

\author[0000-0001-8800-1793]{L.~D.~Anderson}
\affiliation{Department of Physics and Astronomy, West Virginia University, Morgantown, WV 26506}
\affiliation{Adjunct Astronomer at the Green Bank Observatory, P.O. Box 2, Green Bank, WV 24944}
\affiliation{Center for Gravitational Waves and Cosmology, West Virginia University, Chestnut Ridge Research Building, Morgantown, WV 26505}

\author[0000-0001-8061-216X]{Matteo~Luisi}
\affiliation{Department of Physics, Westminster College, 319 S Market St., New Wilmington, PA 16172}
\affiliation{Center for Gravitational Waves and Cosmology, West Virginia University, Chestnut Ridge Research Building, Morgantown, WV 26505}

\author[0000-0002-1311-8839]{B. Liu}
\affiliation{CAS Key Laboratory of FAST, National Astronomical Observatories, Chinese Academy of Sciences, Beijing 100101, People's Republic of China}

\author[0000-0002-4727-7619]{Dylan~J.~Linville}
\affiliation{Department of Physics and Astronomy, West Virginia University, Morgantown, WV 26506}
\affiliation{Center for Gravitational Waves and Cosmology, West Virginia University, Chestnut Ridge Research Building, Morgantown, WV 26505}

\author[0000-0002-8109-2642]{Robert A. Benjamin}
\affil{University of Wisconsin-Whitewater, 800 W. Main St, Whitewater, WI 53190, USA}

\author[0000-0002-5119-4808]{Natasha Hurley-Walker}
\affiliation{International Centre for Radio Astronomy Research, Curtin University, Bentley, WA 6102, Australia}

\author[0000-0003-2730-957X]{N.~M.~McClure-Griffiths}
\affiliation{Research School of Astronomy \& Astrophysics, Australian National University, Canberra 2600 ACT Australia}

\author[0000-0002-2250-730X]{Catherine Zucker}
\affiliation{Harvard Astronomy, Harvard-Smithsonian Center for Astrophysics, 60 Garden St, Cambridge, MA, 02138, USA}


\correspondingauthor{L.D.~Anderson}
\email{loren.anderson@mail.wvu.edu}

\begin{abstract}
The Galactic center lobe (GCL) is an object $\sim\!1\degree$ across that is located north of the Galactic center.  In the mid-infrared (MIR) the GCL appears as two 8.0\,\micron\ filaments between which is strong 24\,\micron\ and radio continuum emission.  Due to its morphology and location in the sky, previous authors have argued that the GCL is located in the Galactic center region, created by outflows from star formation  or by activity of the central black hole Sagittarius~A$^*$.  In an associated paper (Hurley-Walker et al., 2024, in press),
low-frequency radio emission indicates that the GCL must instead lie foreground to the Galactic center.
If the GCL is foreground to the Galactic center, it is likely to be a type of object common throughout the Galactic disk; we here investigate whether its properties are similar to those of Galactic \hii\ regions.  
We find that the GCL's MIR morphology, MIR flux densities, dust temperatures, and radio recombination line (RRL) properties as traced by the GBT Diffuse Ionized Gas Survey (GDIGS) are consistent with those of known Galactic \hii\ regions, although the derived electron temperature is low.  We search for the ionizing source(s) of the possible \hii\ region and identify a stellar cluster candidate (Camargo \#1092/Ryu \& Lee \#532) and a cluster of young stellar objects (SPICY G359.3+0.3) whose members have Gaia parallaxes distances of $1.7\pm 0.4$\,\kpc.
Taken together, the results of our companion paper and those shown here suggest that the GCL has properties consistent with those of an \hii\ region located $\sim\!2\,\kpc$ from the Sun.
\end{abstract}


\section{Introduction}

The Galactic center is an extreme environment, with intense (but inefficient) star formation and a $4\times 10^6\,\msun$ black hole \citep{gillessen09}, Sagittarius\,A$^*$ (``Sgr\,A$^*$'').  Although there is evidence for past accretion in the form of the
``Fermi bubbles'' \citep{su10}, bipolar $\gamma$-ray structures that extend some 50\degree\ above and below the plane, Sgr\,A$^*$ does not appear to be accreting at present.


In the direction of the Galactic center is a structure known as the ``Galactic Center Lobe'' (GCL), sometimes called the ``Galactic Center Omega Lobe.'' Previous authors have suggested that the GCL may be the result of past activity of Sgr\,A$^*$.
The GCL is found north of the Galactic center and is about $1\degree$ across.  At mid-infrared (MIR) wavelengths, the GCL appears as two 8.0\,\micron\ filaments, with the region between them filled with 24\,\micron\ emission. Thermal radio emission from plasma also fills the space between the two 8.0\,\micron\ filaments 
(Figure~\ref{fig:gc}).
\citet{sofue84} first detected the GCL in 10\,\ghz\ continuum data \citep[see also][]{sofue85}.  Their western filament
corresponds to that discussed here, but at Galactic latitudes
$b\lsim0.5\degree$ their eastern filament bends toward more positive
latitudes than what we identify as the GCL here.


Numerous explanations have been explored for the origin of the GCL.  \citet{law09} and \citet{law10} favored an explaination of the GCL being powered by Galactic center star formation; \citet{law10} estimated that three large star clusters in the Galactic center region could account for the ionization of the GCL.
Multiple studies have explored the idea that the GCL results from twisting of poloidal magnetic field lines by Galactic rotation \citep{sofue85, uchida85, uchida90}.  \citet{larosa85, kassim86, bland-hawthorn03} suggested that the GCL is related to activity of Sgr\,A$^*$.
\citet{heywood19} detected the GCL in MeerKAT 1.28\,\ghz\ data, 
and interpreted its emission as arising from an explosive, energetic event near Sgr~A$^*$ that occurred a few million years ago.  There is also X-ray emission seen toward the northern and southern features that further supports the explosion hypothesis \citep{nakashima13,ponti19}.

Complicating these scenarios is uncertainty about the distance to the GCL.  Although it is often taken for granted that objects seen in the direction of the Galactic center are actually in the Galactic center region 8.2\,\kpc from the Sun \citep{abuter19}, recent studies have cast doubt on that assumption for the GCL.  \citet{nagoshi19} argued based on radio recombination line (RRL) emission that the velocities of the detected thermal ionized gas in the GCL are inconsistent with Galactic rotation of an object in the Galactic center.  \cite{tsuboi20} noted that a section of the GCL is correlated with optical extinction feature while a higher latitude section ($b > 0.8\degree$) is seen in H$\alpha$ emission; these results argue for the GCL being foreground to the Galactic center.   
In a companion paper (``Paper~I''; Hurley-Walker et al., 2024, in press), we argue that the GCL is foreground to the Galactic center based on low-frequency radio observations where it is seen in absorption \citep[see also][where this low-radio-frequency absorption signal was first detected]{brogan03}; this signal is best explained by thermal free-free absorption typical of \hii\ regions.

If the GCL is foreground to the Galactic center, and therefore not caused by activity therein, it is likely of a class of object found throughout the Galactic disk.  In an effort to place the GCL in context, we here compare its mid-infrared and radio properties against those of Galactic \hii\ regions.

\section{Data}
\begin{figure}
    \centering
    \includegraphics[width=2.76in]{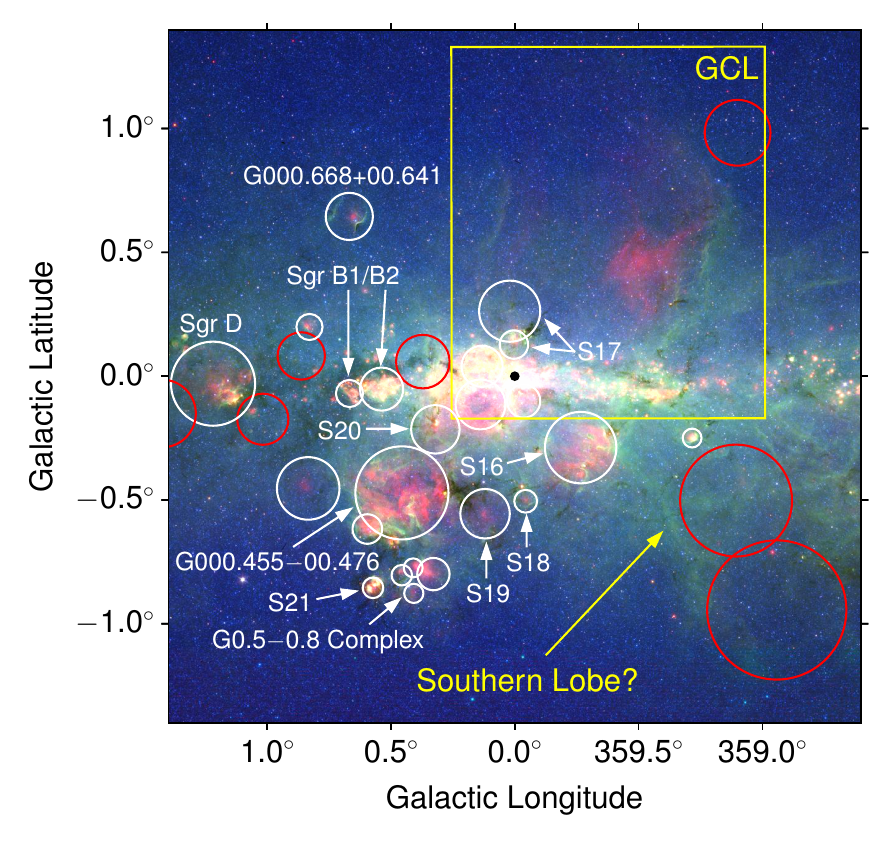} \vskip -8pt
    \includegraphics[width=2.76in]{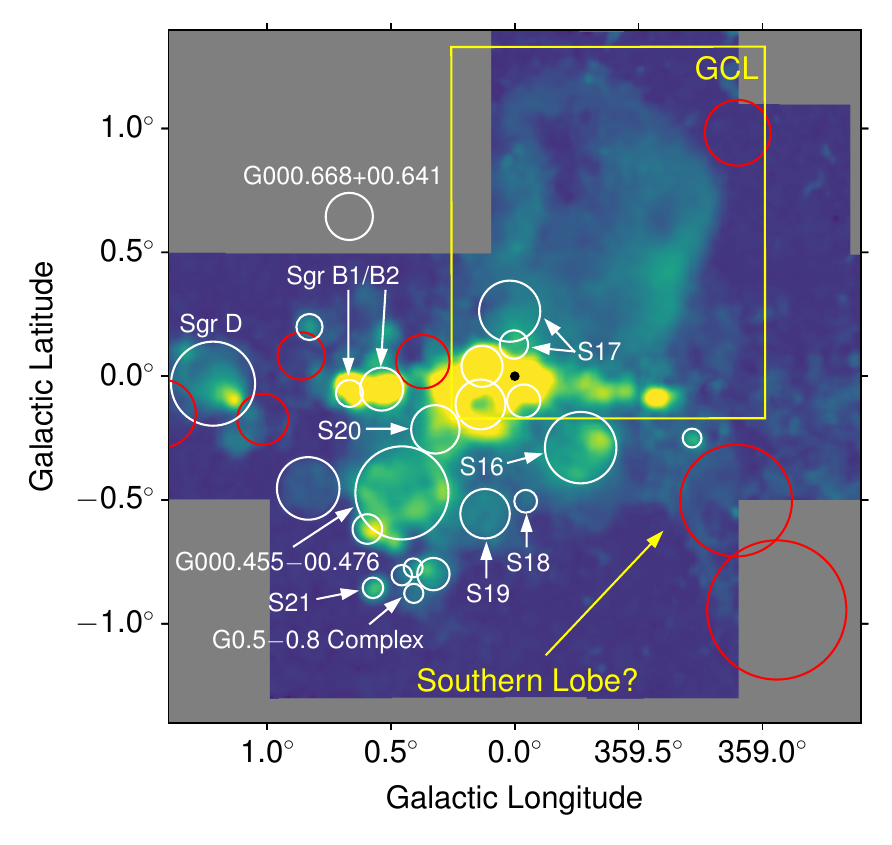} \vskip -8pt
    \includegraphics[width=2.3in]{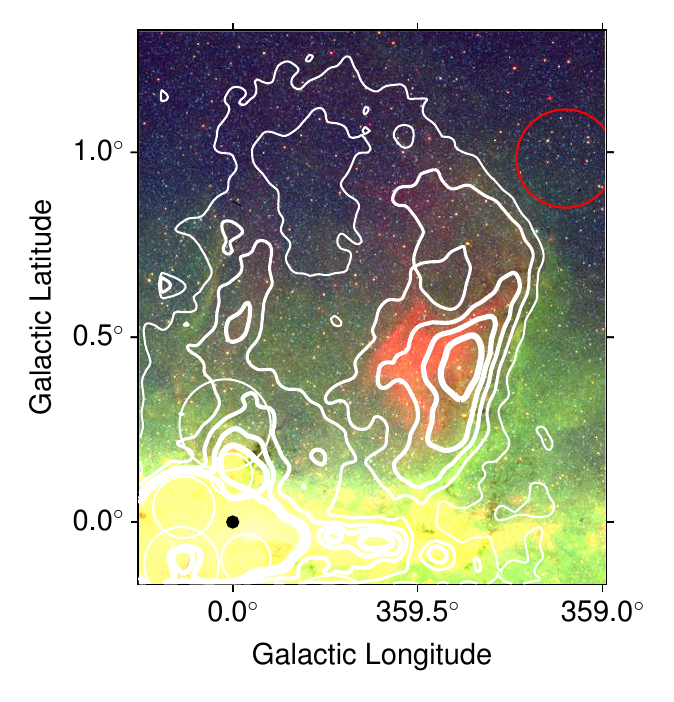} \vskip -8pt
\caption{Top: Spitzer view of the Galactic center region, with MIPSGAL 24\,\micron\ data (red), GLIMPSE 8.0\,\micron\ data (green), and GLIMPSE 3.6\,\micron\ data (blue). Middle: GDIGS integrated Hn$\alpha$ RRL intensity, integrated over $\pm20\,\kms$ (Section~\ref{sec:rrls}).  
Bottom: Inset of the GCL in Spitzer data with white contours of GDIGS emission at levels from 0.5 to 2.5\,\K\kms\ in increments of 0.5\,\K\kms.   
White circles in all panels enclose \hii\ regions from the WISE Catalog of Galactic \hii\ Regions \citep{anderson14} and red circles enclose known supernova remnants \citep{green06}.  The black dots show the location of Sgr~A$^*$. 
The GCL has stronger MIR and RRL emission on its western edge compared to its eastern edge and has weaker emission toward the north. 
    \label{fig:gc}}
\end{figure}

We use data from a variety of sources to investigate the nature of the GCL.  We show overview images of the Galactic center and the GCL in Figure~\ref{fig:gc}.

\subsection{MIR data}
We use 8.0\,\micron\ data from the Spitzer GLIMPSE survey \citep{benjamin03, churchwell09} and 24\,\micron\ data from the MIPSGAL survey \citep{carey09} (see top and bottom panels of Figure~\ref{fig:gc}).  The 8.0\,\micron\ emission is largely from polycyclic aromatic hydrocarbons (PAHs), which fluoresce in the presence of soft ultra-violet ($\sim\!5\,$eV) radiation \citep{voit92}.
At 24\,\micron, most Galactic emission arises from small dust grains irradiated by massive stars.

\subsection{GBT RRL data}
\begin{figure}
    \centering
    \includegraphics[width=3in]{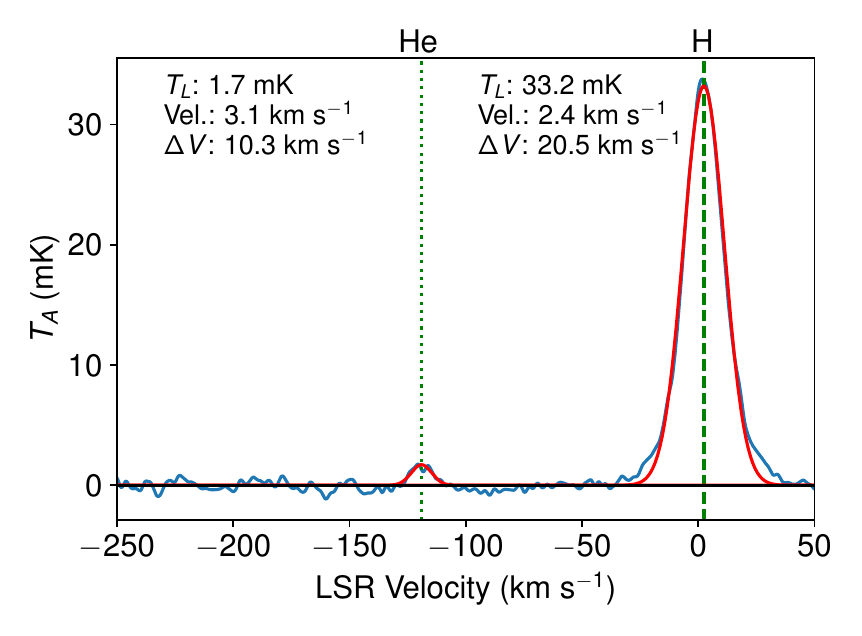}
    \caption{GDIGS RRL emission integrated over the GCL.  We fit Gaussian line profiles and mark the positions of the hydrogen and helium RRLs.  The helium RRL velocity is shifted by $-122.15\,\kms$ from that of hydrogen due to the mass difference between helium and hydrogen.}
    \label{fig:gdigs_gcl}
\end{figure}

We also use data from the Green Bank Telescope (GBT) Diffuse Ionized Gas Survey \citep[GDIGS;][see middle panel of Figure~\ref{fig:gc}]{anderson21}. GDIGS is a fully-sampled C-band (4-8\,\ghz) RRL survey of the inner Galactic plane using the GBT.  
GDIGS simultaneously measures the emission from 22 Hn$\alpha$ transitions, ranging from $n=95$ to 117, and we average the 15 usable RRLs
to increase the signal to noise ratio \citep[see][]{luisi18}. 
The GDIGS maps have $\sim2.65\arcmin$ spatial resolution and 0.5\kms\ velocity resolution.  We show the integrated GDIGS spectrum of the GCL in Figure~\ref{fig:gdigs_gcl}, together with the fits to the hydrogen and helium RRLs.  We create this spectrum by integrating over the entirety of the GCL.

\section{Comparison to Known Galactic HII Regions}
In Paper~I we argue that the GCL is foreground to the Galactic center and is thus unconnected to star formation or black hole activity therein. 
Given that the GCL's radio and MIR morphology are similar to those of Galactic \hii\ regions, in this Section we compare the properties of the GCL to those of known Galactic \hii\ regions.  For this comparison, we use data from the WISE Catalog of Galactic \hii\ Regions \citep[hereafter the ``WISE Catalog'';][]{anderson14}, which contains all known Galactic \hii\ regions (it is updated continuously as new results are published).

\begin{figure*}[!ht]
    \setlength\tabcolsep{0pt}
    \centering
    \begin{tabular}{cccc}
    \includegraphics[width=1.3in]{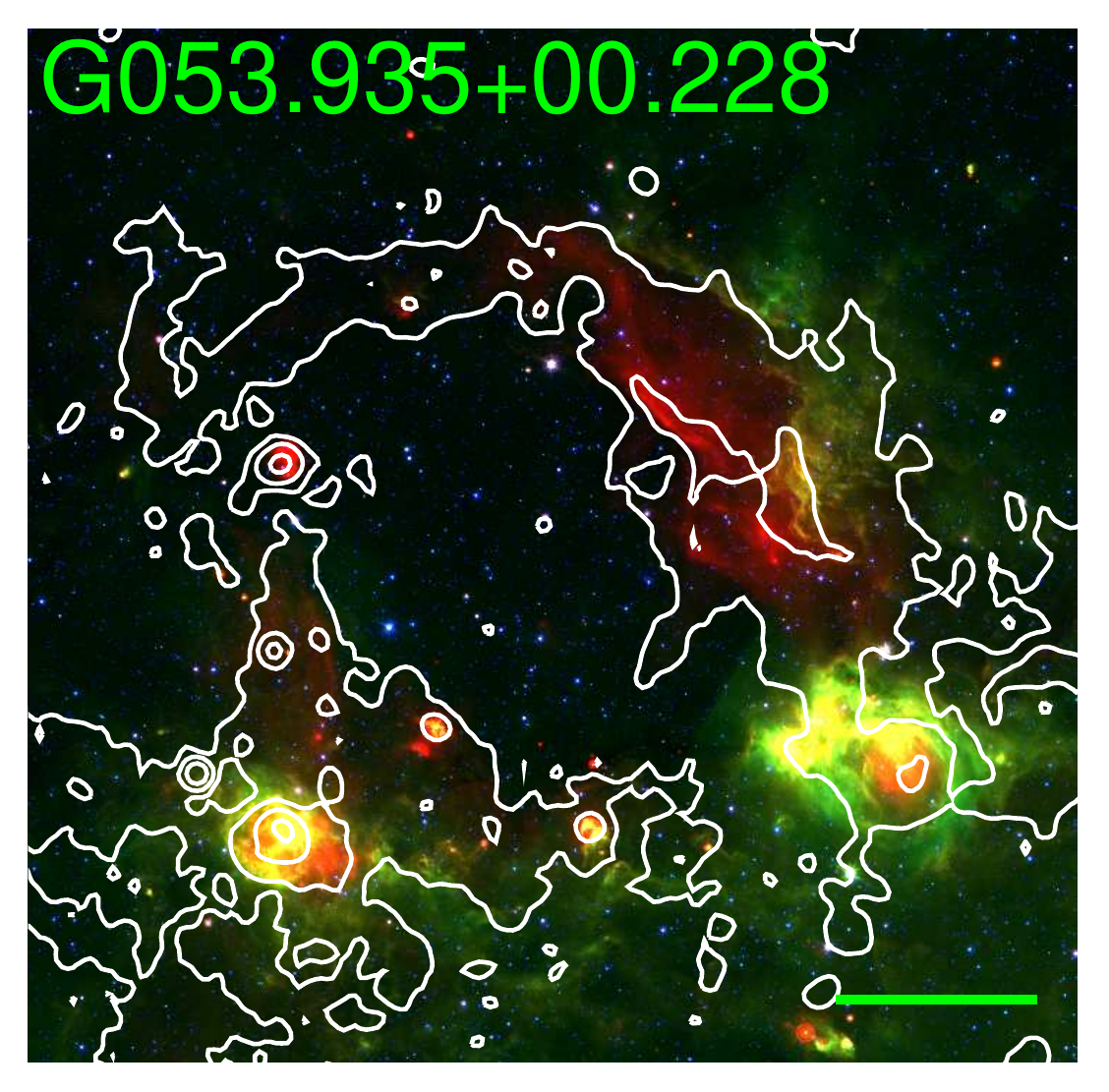} &
    \includegraphics[width=1.3in]{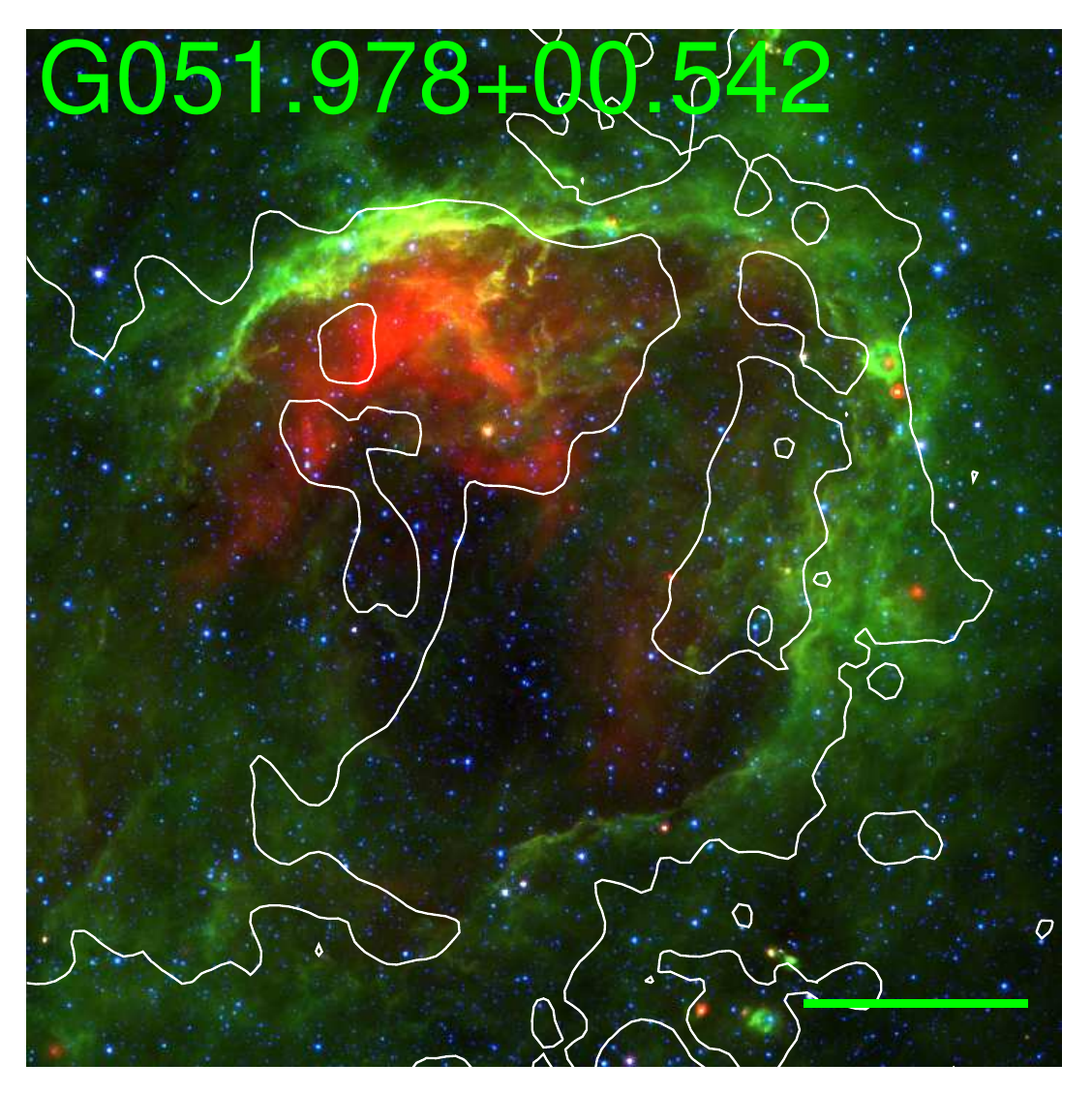} &
    \includegraphics[width=1.3in]{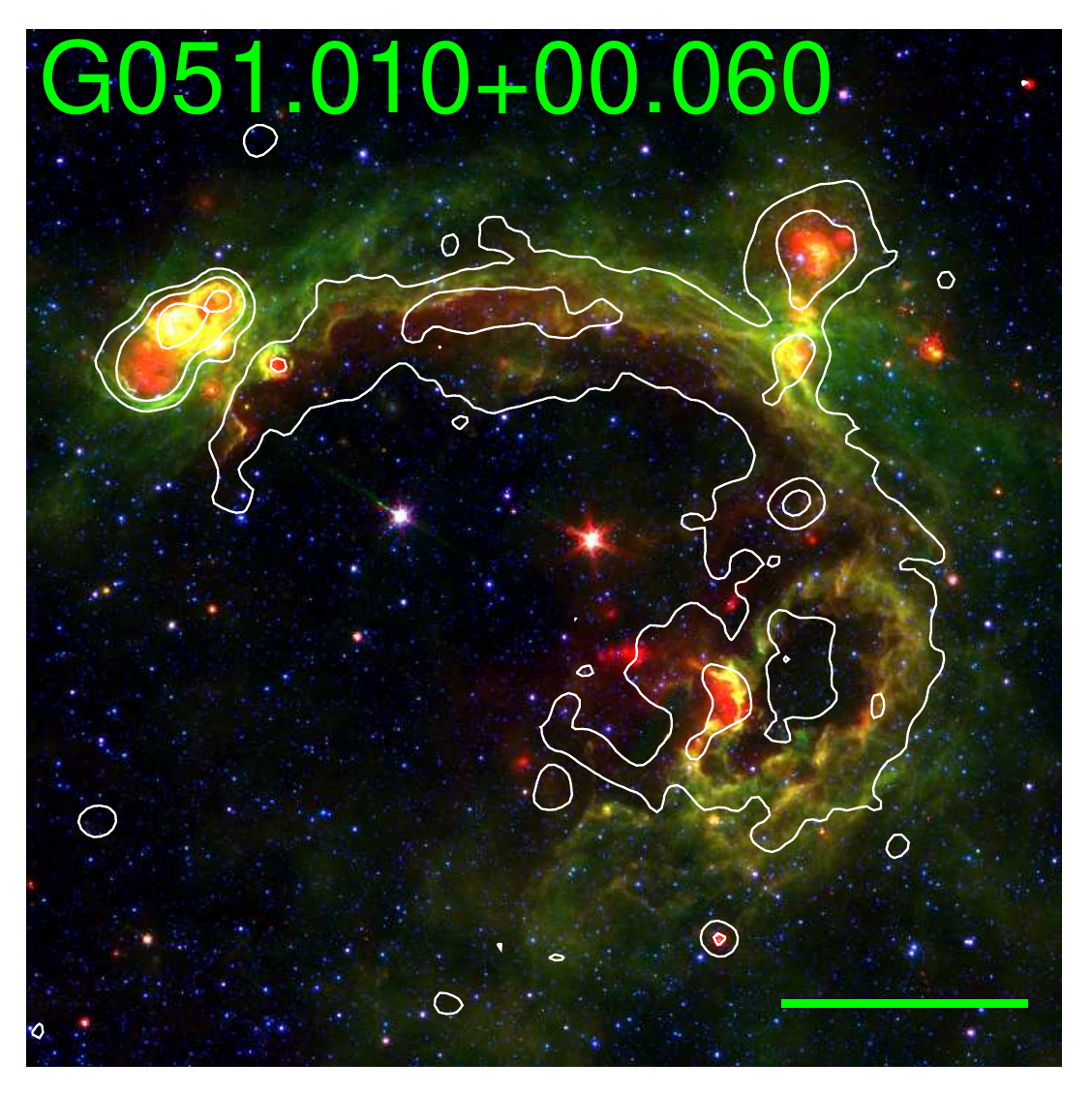} & \phantom{d}
    \multirow{2}*[+1.135in]{\includegraphics[width=2.635in]{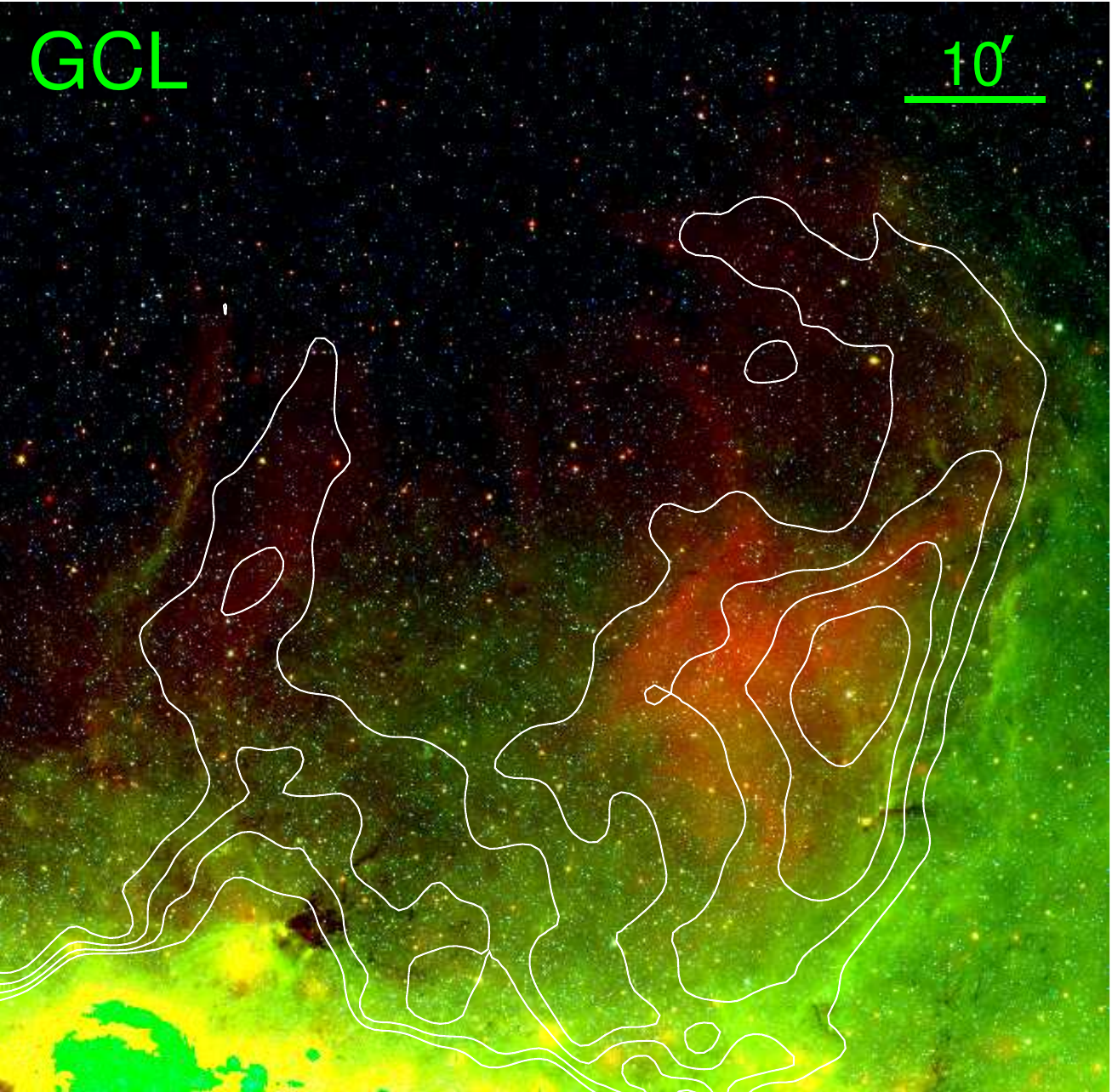}}
\\  \includegraphics[width=1.3in]{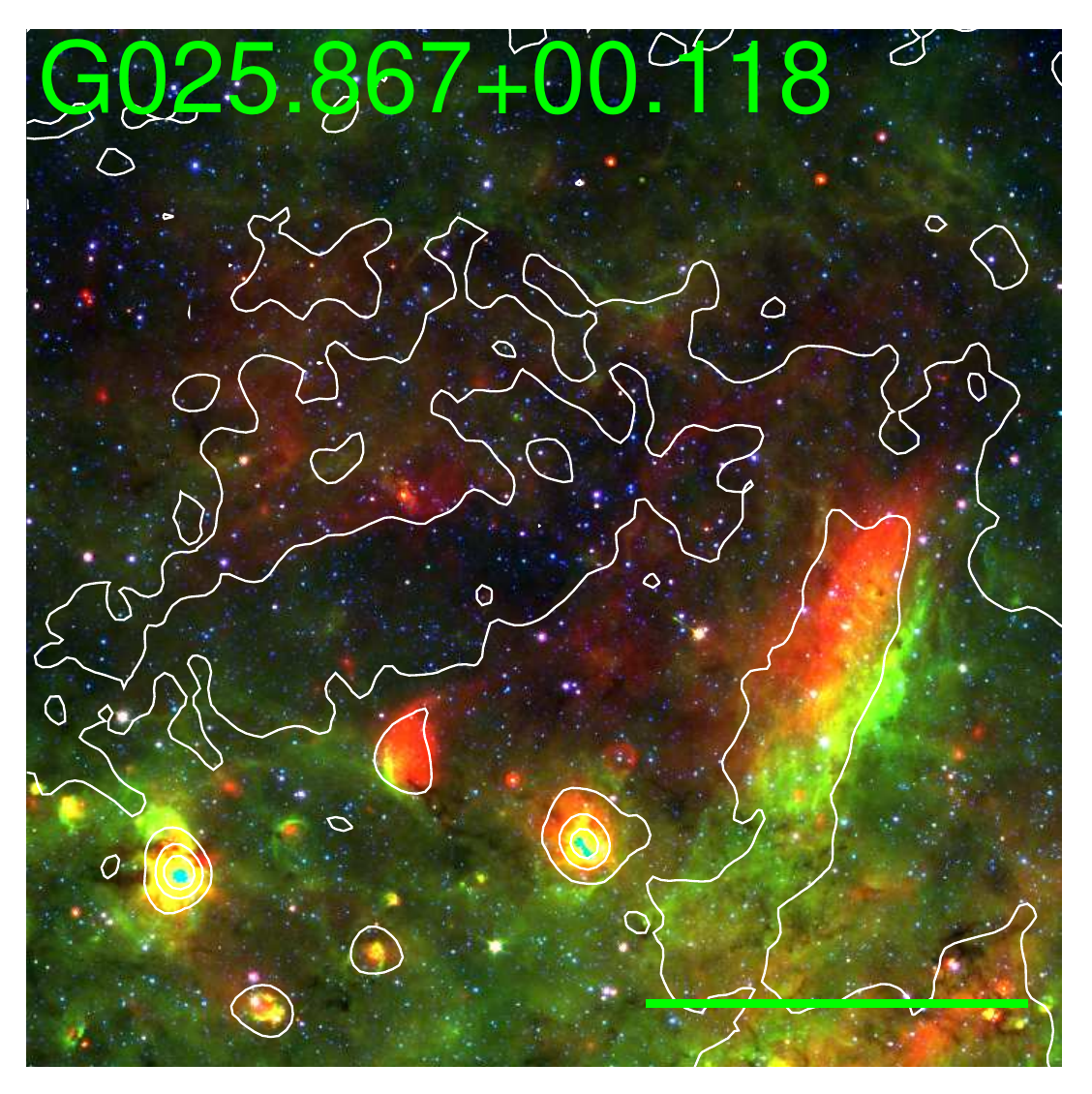} &
    \includegraphics[width=1.3in]{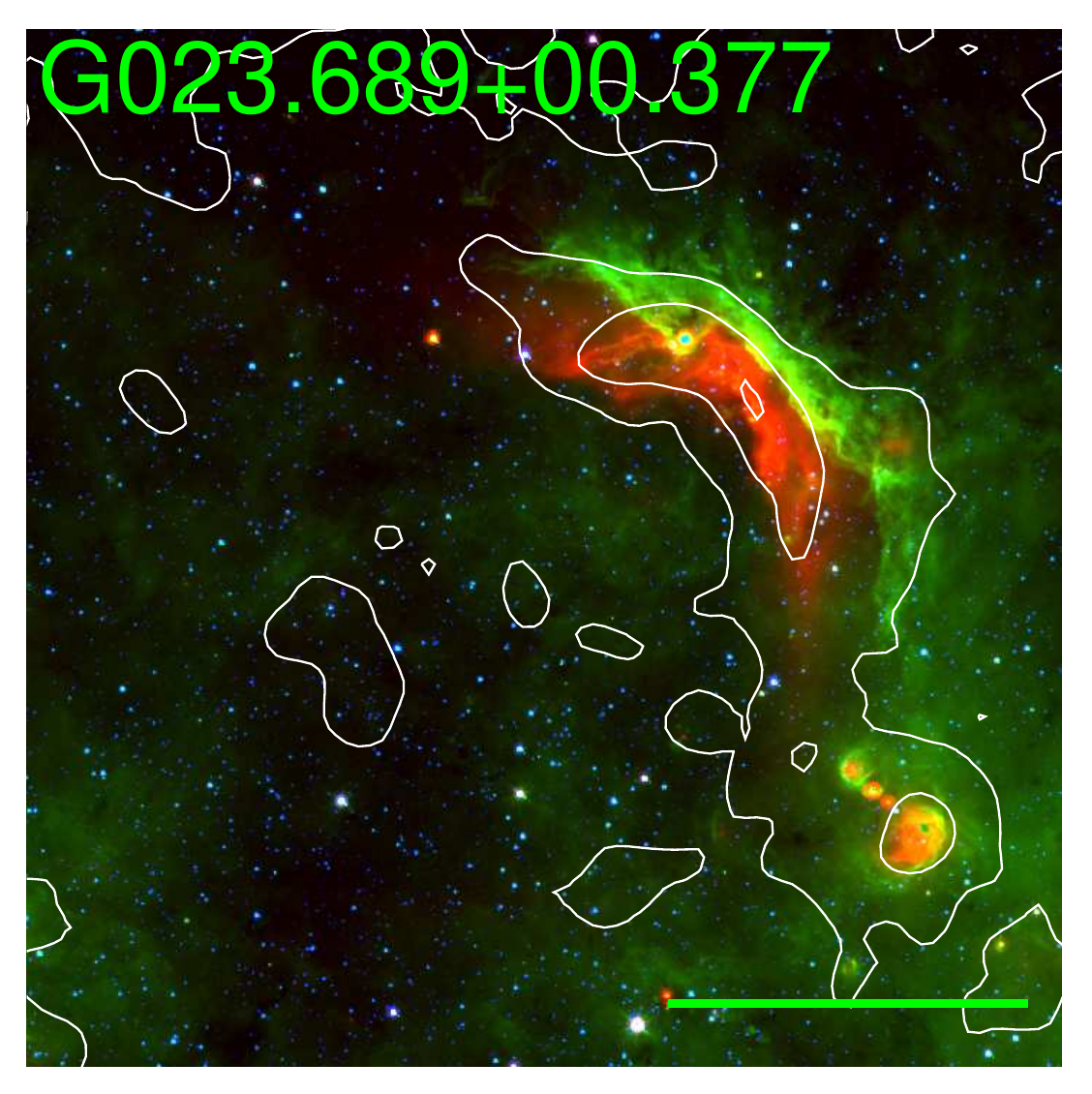} &
    \includegraphics[width=1.3in]{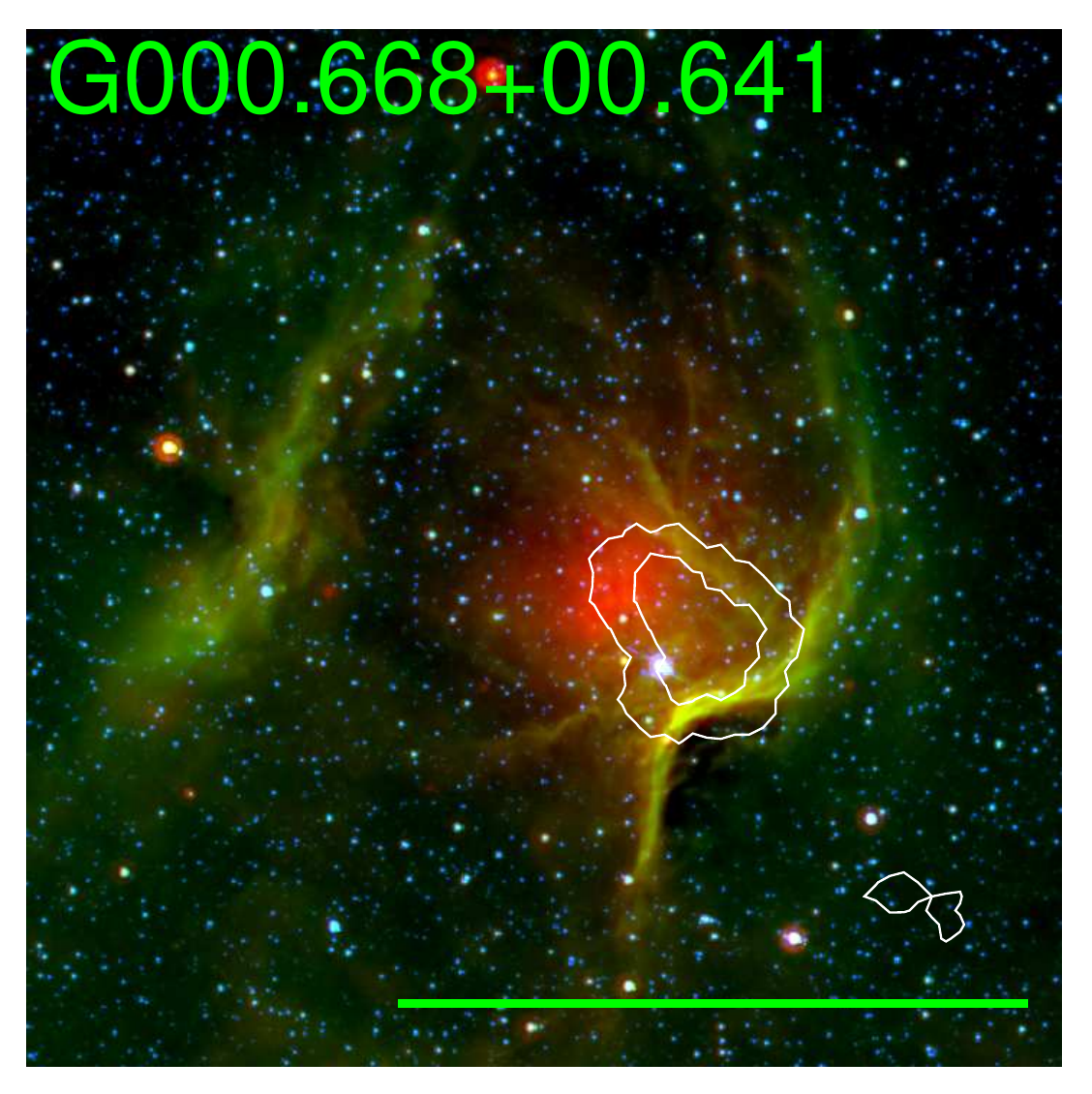}
    \end{tabular}
    \caption{Smaller panels show Spitzer MIR images of first Galactic quadrant \hii\ regions that have a similar morphologies to that of the GCL. 
The larger panel is the GCL with a different color scaling than that of Figure~\ref{fig:gc}.  The colors are the same as in the top and bottom panels of Figure~\ref{fig:gc}: MIPSGAL 24\,\micron\ data in red, GLIMPSE 8.0\,\micron\ data in green, and GLIMPSE 3.6\,\micron\ data in blue.  Contours in the smaller panels are of $\sim\!20\cm$ radio continuum emission: 
NVSS data \citep{condon98} for G000.668+00.641 and 
VLA Galactic Plane Survey data \cite[VGPS;][]{stil06} for the others.  For the larger GCL panel, contours are again of GDIGS RRL emission.  Scale bars in all panels have lengths of $10\arcmin$.
 In all cases the 8.0\,\micron\ emission defines the photodissociaton regions and is not continuous. The 24\,\micron\ and radio continuum data are patchy and are strongest interior to the brightest 8.0\,\micron\ emission. The GCL morphology is similar to that of angularly-large Galactic \hii\ regions. \label{fig:examples}}
\end{figure*}

\subsection{MIR Morphology}
The 8.0\,\micron\ emission of the GCL is mainly comprised of western ($\ell=359-360\degree$) and eastern ($\ell\simeq0\degree$) filaments, with 24\,\micron\ emission filling the space between the filaments (Figure~\ref{fig:gc}).  The western edge of the GCL is stronger in 8.0\,\micron\ emission compared to the eastern edge and the 8.0\,\micron\ emission is absent toward the north.  The strongest 24\,\micron\ emission is found interior to the western edge.  Radio continuum emission fills the area bounded by the 8.0\,\micron\ edges, and has a similar morphology to the 24\,\micron\ emission.


The MIR morphology of the GCL is characteristic of Galactic \hii\ regions. \citet{anderson14} found that all known Galactic \hii\ regions have the same basic MIR morphology seen for the GCL: $\sim\!10\,\micron$ emission surrounding $\sim\!20\,\micron$ emission that is spatially coincident with the thermally-emitting ionized gas.  
No other known class of astrophysical object has the same morphology. Supernova remnants, for example, can have $\sim\!20\,\micron$ emission \citep[see][for a review]{pinheiro11} but lack $\sim\!10\,\micron$ emission.  Planetary nebulae can have both $\sim\!10$ and $20\,\micron$ emission, but due to their angular sizes, this emission is rarely resolved.  The same issue can apply to luminous blue variable stars.  The morphology therefore indicates the presence of $\sim\!10^4\,\K$ plasma created by a central radiation source.

In Figure~\ref{fig:examples} we show examples of angularly-large first Galactic quadrant \hii\ regions from the WISE Catalog that have MIR morphologies similar to that of the GCL.  
As for the GCL, the 8.0\,\micron\ emission of the example \hii\ regions is not continuous and is intense in some directions and not detected in others.  The 24\,\micron\ emission for the example regions is patchy and is strongest interior to the most intense 8.0\,\micron\ emission.  The radio continuum emission is found co-spatial with the 24\,\micron\ emission.  All of these features are similar to those of the GCL.

\subsection{MIR Flux Densities}
Galactic \hii\ regions have characteristic MIR flux density ratios \citep[see][]{makai17}.  We measure the 8.0 and 24\,\micron\ flux densities of the GCL using aperture photometry with multiple background apertures. We take the standard deviation of the derived flux densities from the variations from different background apertures as the uncertainty in the measurements \citep[see][]{makai17}.  Due to bright and variable background emission in the direction of the GCL, the uncertainties in the derived flux densities are large ($\sim\!100\%$).

\begin{figure*}
    \centering
    \includegraphics[width=2.8in]{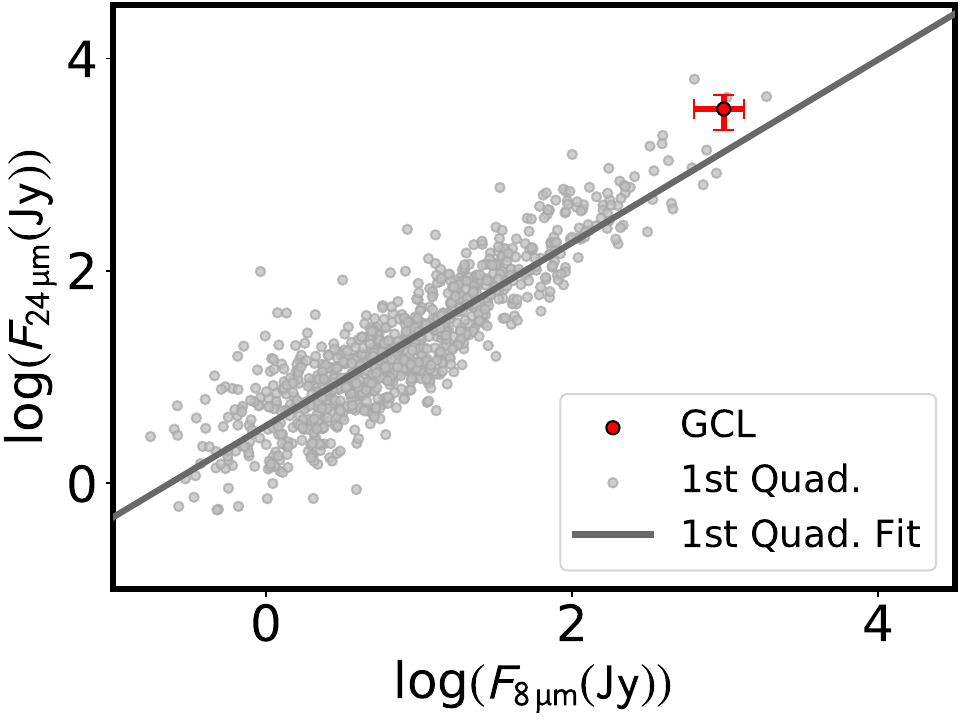}
    \includegraphics[width=2.8in]{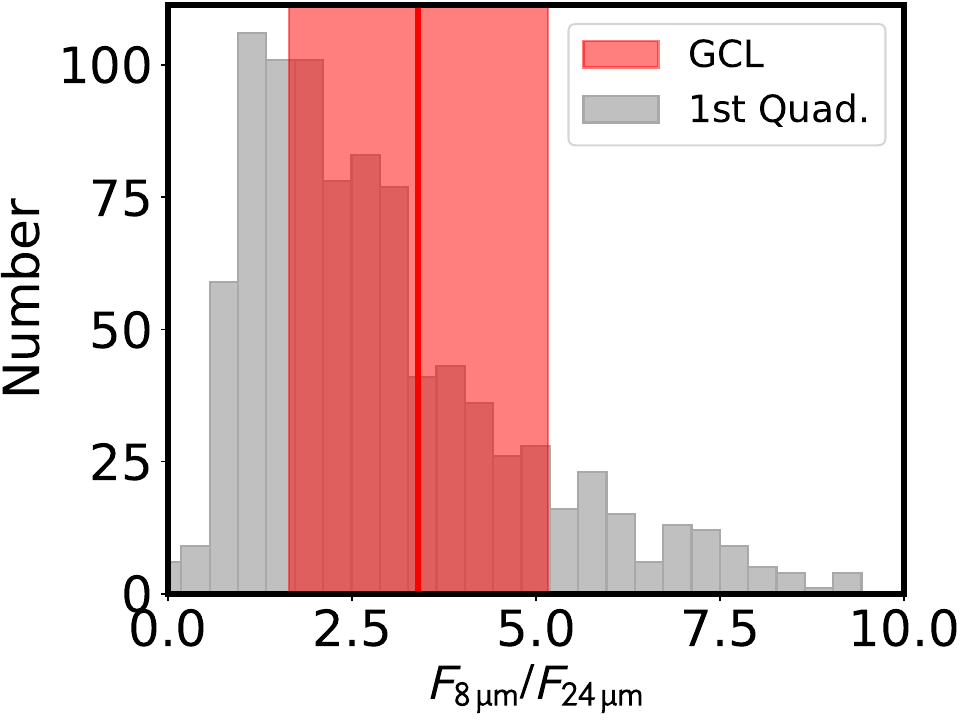}
    \caption{Left: Scatter plot of 8.0 and 24\,\micron\ flux densities of the GCL and the first-quadrant \hii\ region sample of \citet{makai17}, with their best fit line.  The GCL is brighter than nearly all \hii\ regions in the Galaxy.  Right: Histogram of the 8-to-24\,\micron\ flux density ratios for the same \hii\ region sample. The GCL's 8-to-24\,\micron\ flux density ratio is aerage for Galactic \hii\ regions.}
    \label{fig:fluxes}
\end{figure*}

In Figure~\ref{fig:fluxes}, we compare the derived flux density values of the GCL with those of the large sample of first-quadrant WISE Catalog \hii\ regions from \citet{makai17}.  The GCL is brighter than nearly all first quadrant Galactic \hii\ regions, but its 8-to-24\,\micron\ flux density ratio is not unusual.  For example, about 30\% of the \citet{makai17} sample have 8-to-24\,\micron\ flux density ratios greater than that of the GCL.

\subsection{Dust Temperature}
The dust temperature of the GCL is similar to that of \hii\ regions.  The IRAS-derived dust temperature of the western edge of the GCL is $T_d = 27\pm 1\,\K$ \citep{gautier84}.  \citet{anderson12b} found that the photodissociation regions of a sample of ``bubble'' \hii\ regions have average dust temperatures of 26\,\K, and furthermore that there is little variation in dust temperature in their sample. 

\subsection{Size\label{sec:sizes}}
The GCL has a diameter of $\sim1\degree$, measured as the distance between the eastern and western photodissociation regions.  The largest \hii\ regions in the Galaxy are $\sim\!100\,\pc$ in diameter \citep[e.g.][]{bania12, anderson18}.  The size of an \hii\ region depends on its age, since \hii\ regions expand as they age, and also the luminosity of its ionizing sources, since the expansion is faster for more luminous sources.  If located in the Galactic center, the large angular size of the GCL would imply a physical size that is larger than any known \hii\ region, which would make its identification as an \hii\ region problematic.

\begin{figure}
    \centering
    \includegraphics[width=3.25in]{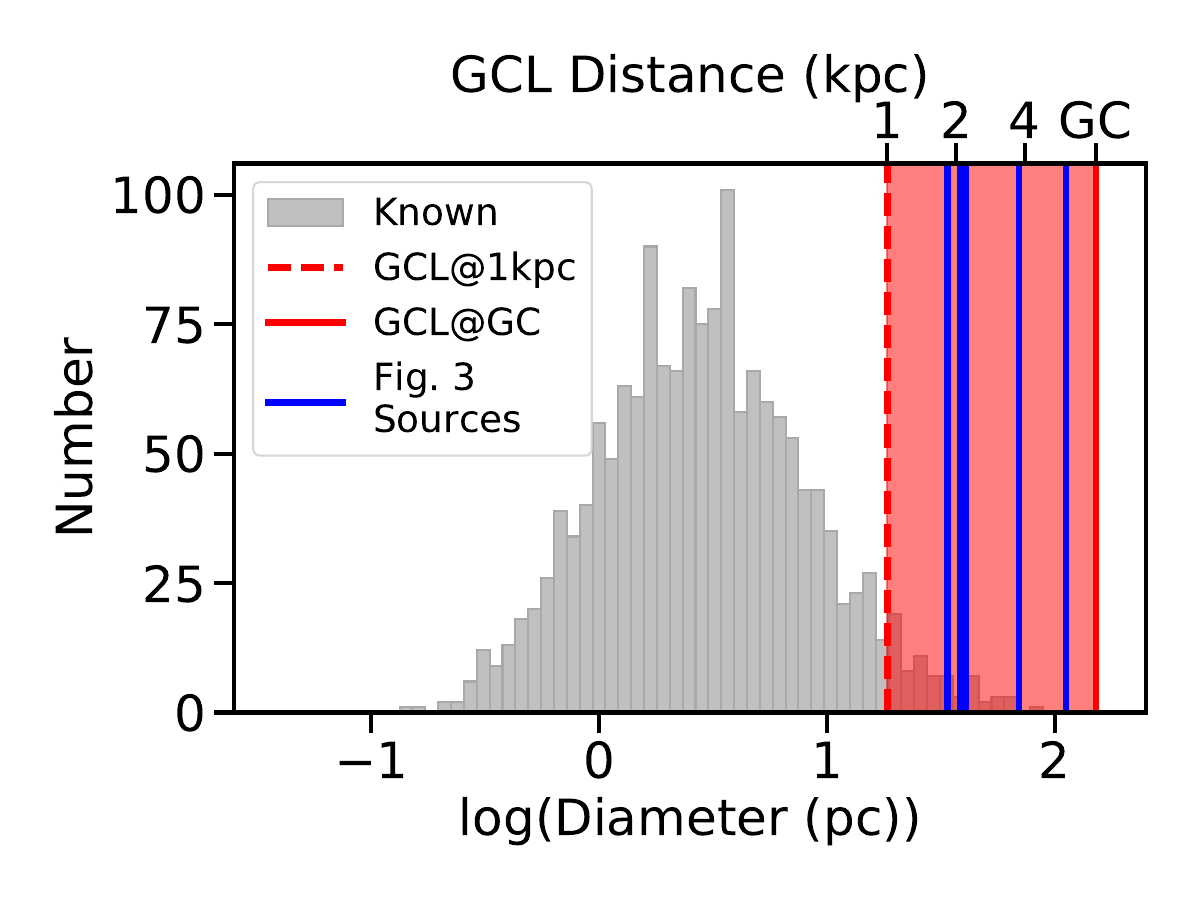}
    \caption{Physical diameter of Galactic \hii\ regions.  The \hii\ region sample shown in gray is that of the WISE Catalog.  The red shaded area shows the diameter of the GCL at heliocentric distances ranging from 1\,\kpc\ to the Galactic center (8.2\,\kpc).  Blue lines show the diameters of the Figure~\ref{fig:examples} \hii\ regions whose MIR emission is similar to that of the GCL and that have known heliocentric distances.  If it is an \hii\ region, the large physical size of the GCL for the range of distances explored is consistent with the sizes of the Figure~\ref{fig:examples} \hii\ regions.
    \label{fig:sizes}}
\end{figure}

In Figure~\ref{fig:sizes} we compare the diameter of the GCL at distances from 1\,\kpc\ to the Galactic center (8.2\,\kpc) to the distribution of all known \hii\ region diameters from the WISE Catalog.   We also plot the diameters for the \hii\ regions from Figure~\ref{fig:examples} that have known heliocentric distances.  The WISE Catalog diameters are also derived from MIR data and so are comparable to that used here for the GCL. 

If the GCL is a foreground \hii\ region, its size would be comparable to other large \hii\ regions in the Galaxy.  Figure~\ref{fig:sizes} shows that the \hii\ regions that have similar morphologies to that of the GCL from Figure~\ref{fig:examples} are all physically large, ranging from $\sim\!30$ to $>\!100\,\pc$ in diameter.  In contrast, the median WISE Catalog \hii\ region diameter is just $\sim\!3\,\pc$.  The GCL at a distance of 2\,\kpc\ \citep[see Paper~I;][]{nagoshi19, tsuboi20} would have a diameter of $\sim\!35\,\pc$ and if located at the distance to the Galactic center would have a diameter of $\sim\,150\,\pc$.  The morphology of the GCL suggests that if it is an \hii\ region its physical size is large no matter its distance.  

\subsection{GDIGS RRL Analyses\label{sec:rrls}}
Using GaussPy+ \citep{riener19} we fit Gaussian components to each GDIGS spaxel that has a signal strength of at least 3$\sigma$, where $\sigma$ is the spectral rms computed individually from the line-free portions at each spaxel, and thereby derive the RRL linewidth, velocity centroid, and intensity across the GCL \citep[see][]{anderson21}.  We use these maps of RRL properties in the analyses below.

\subsubsection{Hydrogen RRL Line Width\label{sec:linewidth}}
Cm-wave RRLs from \hii\ regions have line width contributions from thermal and turbulent broadening, generally with minimal contributions from pressure broadening.  
The thermal broadening of Hydrogen RRLs, $\sigma_{v_{th}}$, is related to the electron temperature via:
\begin{equation}
    \sigma_{v_{th}} = \left(\frac{k_B T_e}{m_p} \right)^{0.5}\,,
\end{equation}
where $k_B$ is the Boltzmann constant and $m_p$ is the proton mass.  For a pure hydrogen plasma the thermal FWHM $\Delta V_{th}$ is
\begin{equation}
    \Delta V_{th} = (8\ln 2)^{0.5} \sigma_{V_{th}} = 21.5 \left(\frac{T_e}{10^4\,\K} \right)^{0.5}\,\kms\,.
    \label{eq:deltav}
\end{equation}

The Hydrogen RRL FWHM values of Galactic \hii\ regions generally fall between $\sim\!20$ and $30\,\kms$ \citep{anderson11} and therefore if the GCL were an \hii\ region we would expect its RRL line width to fall within this range. 
Previous RRL observations by \citet{law09} and \citet{nagoshi19} showed that the GCL has narrow Hydrogen FWHM line widths ranging from $\Delta V \simeq\!10$ to 30\,\kms, depending on the location within the GCL.  \citet{law09} found that the plasma was in LTE, and that the narrow line widths were not caused by stimulated emission.  Narrow line widths of $\sim\! 10\,\kms$ are unusual for Galactic \hii\ regions \citep[see][]{anderson17}, as they imply low electron temperatures, a low amount of turbulence, or both.

\begin{figure}
    \centering
    \includegraphics[width=3in]{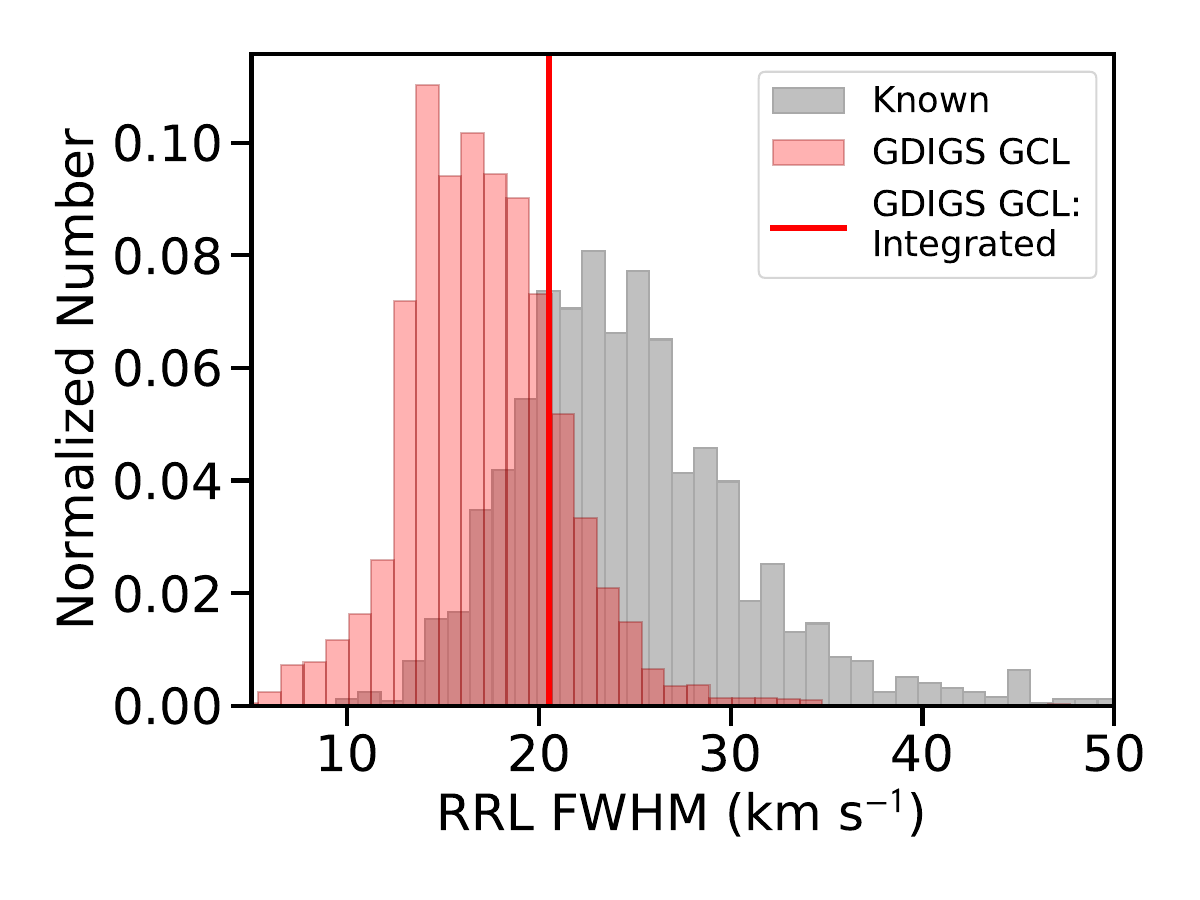}
    \caption{Hydrogen RRL FWHM distribution for Galactic \hii\ regions.  The known population, compiled from the literature in the WISE Catalog, is in gray.  Fits to individual GDIGS spaxels toward the GCL yield FWHM values whose distribution peaks at 15\,\kms, but the FWHM of the GDIGS RRLs integrated over the entire GCL is 20.5\,\kms.  Thus, the integrated FWHM value for the GCL is greater than the peak of the distribution and the integrated GCL FWHM value, which is what one would measure if the GCL had an angular size comparable to the telescope beam, is not unusual for a Galactic \hii\ region. 
    \label{fig:fwhm}}
\end{figure}

In Figure~\ref{fig:fwhm}, we show histograms of the RRL FWHM values for known \hii\ regions in the WISE Catalog and for spaxel-by-spaxel single-Gaussian fits to the GDIGS RRL data of the GCL.  The known \hii\ region FWHM values are not a homogeneous data set, but rather are compiled from pointed observations at frequencies from 5-10\,\ghz using a variety of telescopes.  The average \hii\ region RRL FWHM is near 25\,\kms, and the distribution is roughly normal with a standard deviation of $\sim\!5\,\kms$.  The GCL RRL data have a mean FWHM value near 15\,\kms, and the majority of spaxels have RRL FWHM values between 10 and 25\,\kms.  These results are in line with those from \citet{law09} and \citet{nagoshi19}.  

Figure~\ref{fig:fwhm} shows that the RRL FWHM found from integrating over the GCL, $\Delta V = 20.5\,\kms$ (Figure~\ref{fig:gdigs_gcl}), is similar to those of known \hii\ regions.  
Each known \hii\ region RRL line width in Figure~\ref{fig:fwhm} is from a pointed observation.  Since most \hii\ regions have angular sizes similar to those of the telescope beams used in the observations \citep[e.g.,][]{anderson11}, most FWHM values represent quantities integrated over entire sources.  
We conclude from this analysis that the GCL line width is narrow for an \hii\ region, but not exceptionally so.  The narrow GCL RRL values first reported by \citet{law09} and \citet{nagoshi19} are in part caused by the fact that individual observations were sampling only a small portion of the GCL, and therefore were not broadened by any bulk motions.


Some narrow-line \hii\ regions are known in the literature.  \citet{anderson15b} found four narrow-line RRL sources that have FWHM values $<10\,\kms$; all four have multiple RRL line components and exist in the inner Galaxy.  There was only one such narrow-RRL source in the large sample of RRL \hii\ region observations in \citet{anderson11}, and it too was a multiple RRL source in the inner Galaxy.  
\citet{anderson17} found seven \hii\ regions that have narrow line widths $<10\,\kms$ FWHM.  In contrast to the GCL, the spectra for all seven sources have multiple RRL components. For six of the seven, the narrow component is blended with a broader component.   All seven are located toward the inner Galaxy at low Galactic latitudes.

\subsubsection{Electron Temperatures\label{sec:Te}}

Here, we investigate whether the GCL has derived electron temperature values consistent with those of known Galactic \hii\ regions and with its RRL FWHM values.  Previous studies have found a strong \hii\ region electron temperature gradient with Galactocentric radius ($R_{\rm gal}$): from $T_{\rm e} \approx 5000$\,K near the Galactic center to $T_{\rm e} \approx 12{,}000$\,K at $R_{\rm gal} = 20$\,kpc \citep{quireza06b, balser11,balser16}. 

This $T_{\rm e}$ gradient is thought to be caused by a metallicity gradient.  The electron temperature of an \hii\ region can be used as a proxy for its metallicity. Metals are the primary coolants in \hii\ regions and lower electron temperatures therefore correspond to higher metallicities \citep[e.g.,][]{shaver83}.

Paper~I used archival GBT radio continuum data in combination with GDIGS RRL data to create a map of $T_{\rm e}$ across the GCL.  Assuming that the gas is in local thermodynamic equilibrium (LTE),
\begin{multline} \left( \frac{T_{\rm e}}{\textnormal{K}} \right) = \left \{ 7103.3 \left( \frac{\nu _{\rm L}}{\textnormal{GHz}}\right)^{1.1} \left( \frac{T_{\rm C}}{T_{\rm L}(\textnormal{H}^+)} \right) \right. \\
\left. \left( \frac{\Delta V(\textnormal{H}^+)}{\textnormal{km\,s}^{-1}} \right)^{-1} \left( 1+y^+ \right)^{-1} \right \}^{0.87},\label{eq:etemp}
\end{multline}
where $\nu_{\rm L}$ is the average frequency of the RRL and continuum data, $T_{\rm C}$ is the continuum brightness temperature, $T_{\rm L}(\textnormal{H}^+)$ is the H RRL brightness temperature, $\Delta V
(\textnormal{H}^+)$ is the H RRL FWHM, and $y^+$ is the helium-to-hydrogen ionic abundance ratio,
\begin{equation} y^+ = \frac{T_{\rm L}(^4\textnormal{He}^+)\Delta V (^4\textnormal{He}^+)}{T_{\rm L}(\textnormal{H}^+)\Delta V (\textnormal{H}^+)},
\label{eq:yplus}
\end{equation}
where $T_{\rm L}(^4\textnormal{He}^+)$ is the line temperature of helium and $\Delta V (^4\textnormal{He}^+)$ is the corresponding FWHM line width \citep{Peimbert1992}. Paper~I found that $y^+ = 0.03$ on average over the GCL field.

The derived GCL electron temperatures range from 2{,}000 to 10{,}000\,\K, with 
values of $T_e \simeq 3{,}500$\,K found at locations of the highest RRL intensity; we accept the value of $3{,}500$\K as being representative of that of the GCL \citep[cf.][who argue for a maximum of 3{,}960\,\K]{law08}.  Paper~I estimates 10\% uncertainties on $T_{\rm e}$ due to the radio continuum. This value of the electron temperature would be low for a Galactic \hii\ region and would be particularly low for an \hii\ region near to the Sun. 

\begin{figure}
    \centering
    \includegraphics[width=3.0in]{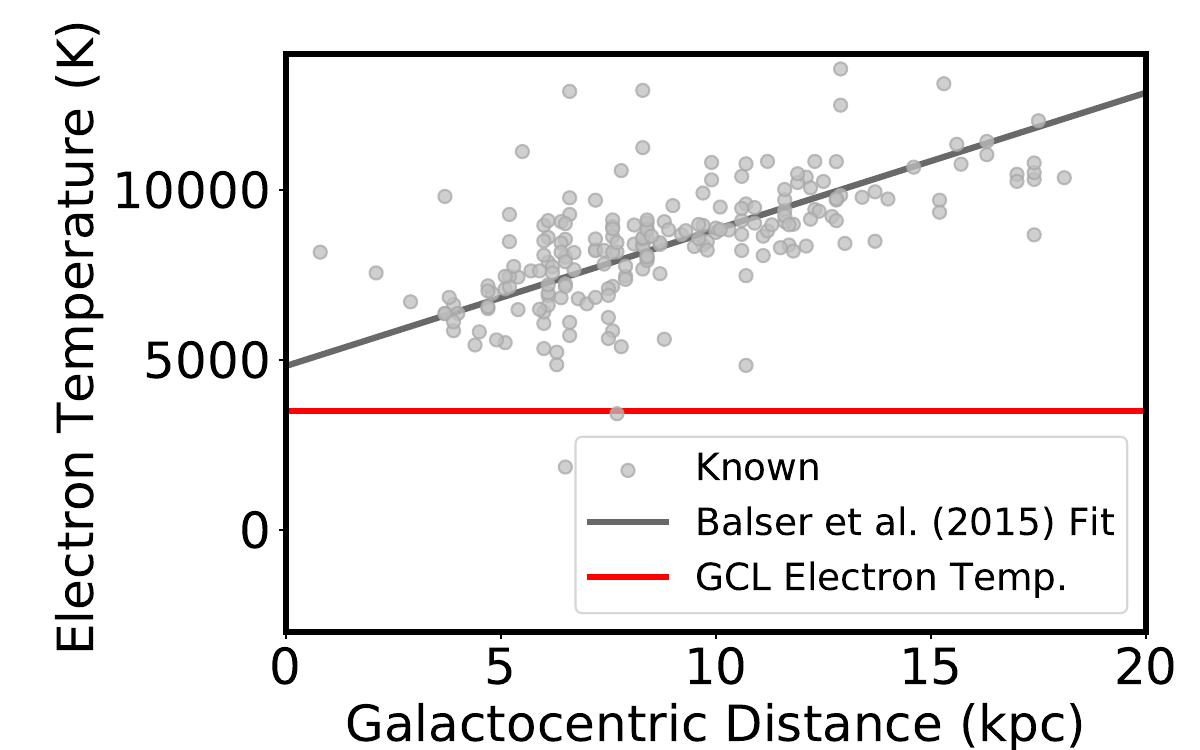}
    \caption{Electron temperatures of \hii\ regions from \citet{balser16}, together with the fit from that paper.  The electron temperature of the GCL, $T_e\simeq 3500\,\K$, is shown as the red horizontal line.  The value of $T_e$ for the GCL is anomalously low for a Galactic \hii\ region, for all Galactocentric radii.
    \label{fig:te_scatter}}
\end{figure}

\begin{figure*}
    \centering
\includegraphics[width=2.2in]{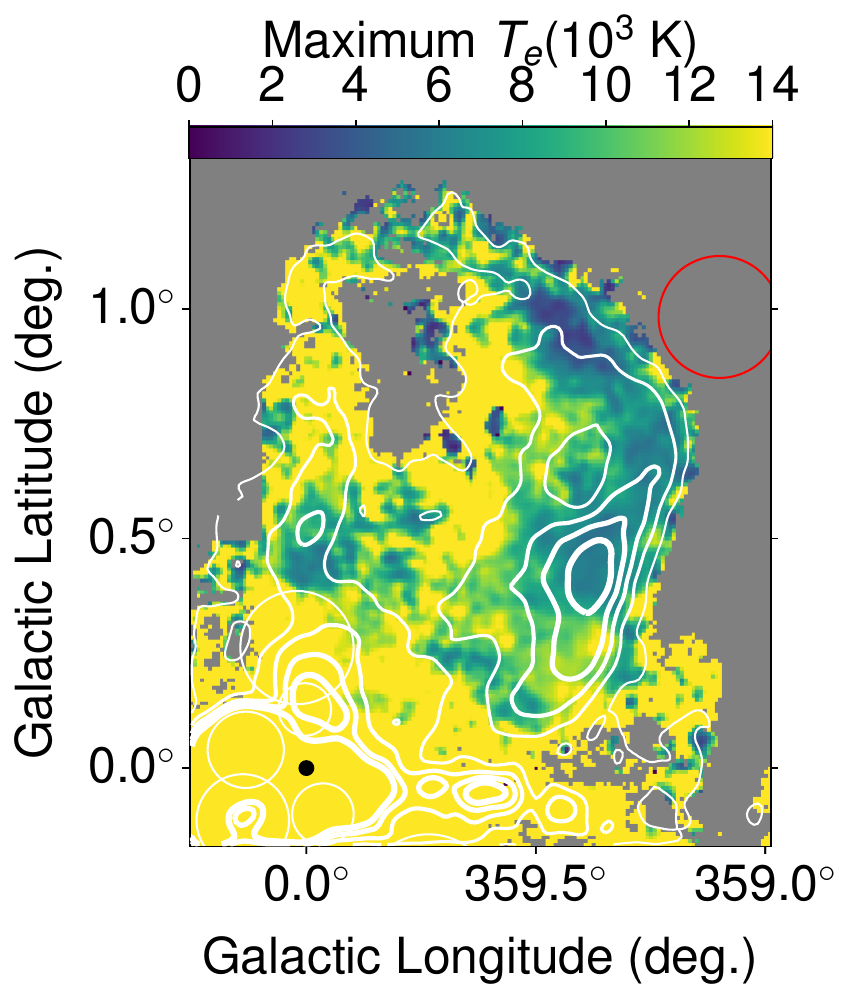}
    \includegraphics[width=2.2in]{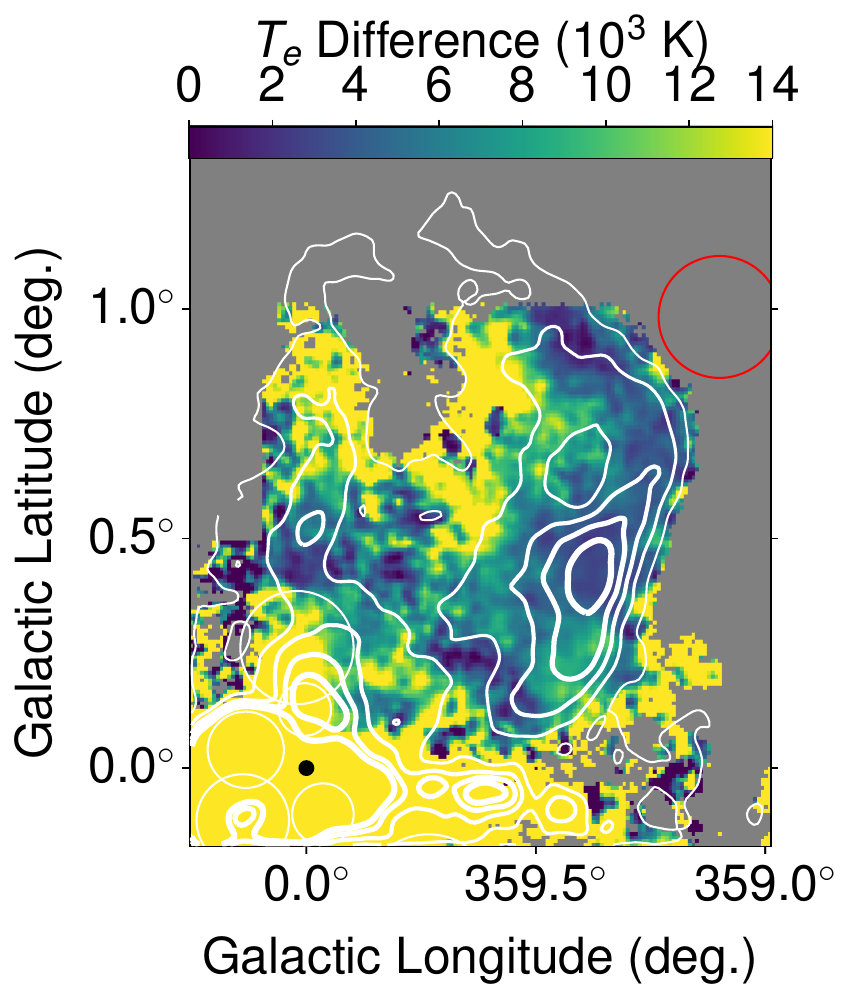}
    \includegraphics[width=2.6in]{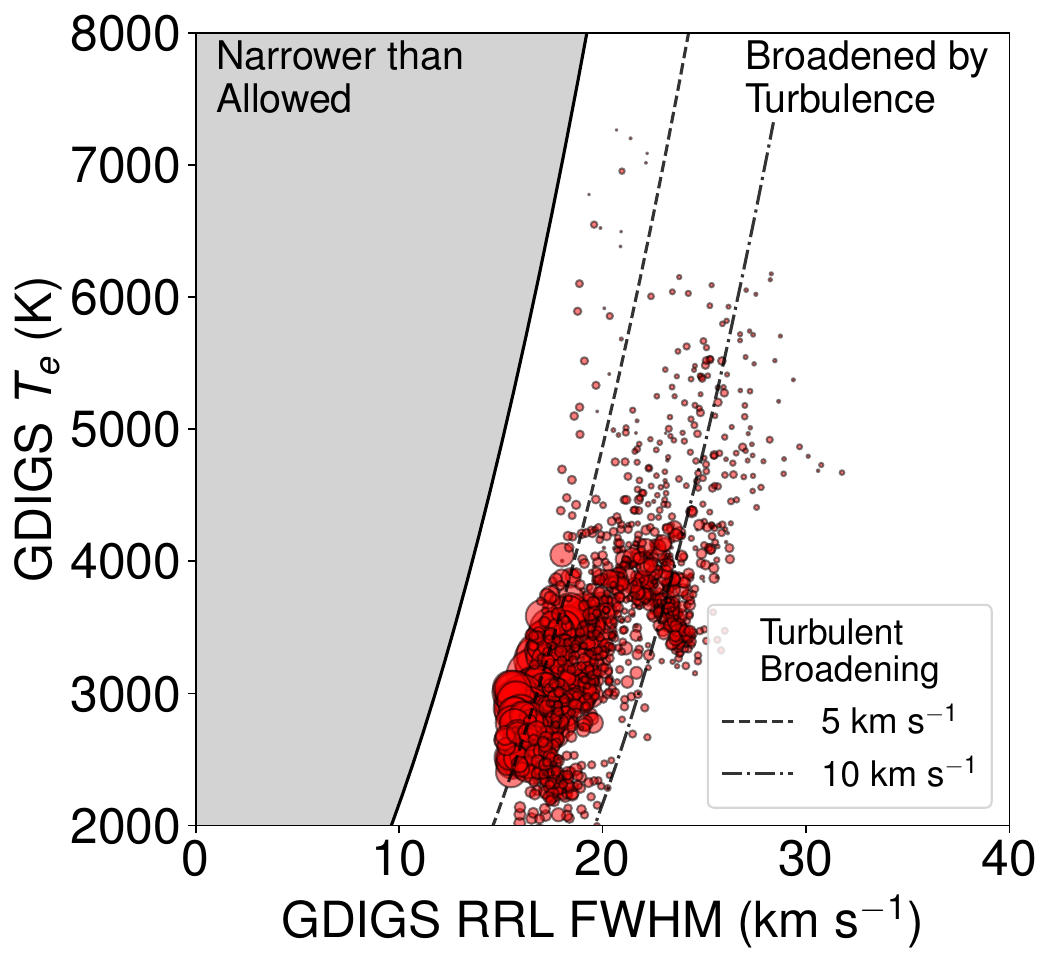}
        \caption{Comparison of the maximum values of $T_e$ allowed from the RRL FWHM values and the derived $T_e$ values themselves.  
    Left: The maximum possible electron temperature values derived from the GDIGS RRL line widths.  A typical maximum electron temperature value at locations of bright RRL emission is $\sim\!7{,}000\,\K$ ($\Delta V \simeq\!18\,\kms$).  
    Middle: The difference between the values of $T_e$ derived in Paper~I and the FWHM-derived map.  The differences are $\sim 2{,}000\,\K$ at the locations of highest RRL intensity, which can be ascribed to turbulent broadening; a value of zero implies that there is no turbulent broadening.  Both panels have the same size as the bottom panel of Figure~\ref{fig:gc}, as well as the same symbols and white contours of GDIGS RRL integrated intensity.  
    Right: Pixel-by-pixel analysis of $T_e$ derived in the two methods.  Symbol sizes are proportional to the GDIGS integrated intensities.  The black curve shows the expected electron temperature for given FWHM values, assuming no other broadening mechanisms (i.e., no turbulence).  Dashed and dot-dashed curves show the expected electron temperature with 5 and 10\,\kms\ of turbulent broadening, respectively. The narrow line widths are allowed for the derived electron temperature values, although the turbulent broadening must be only $\sim\,5$ to 10\,\kms.}
    \label{fig:te_comparison}
\end{figure*}

We compare the $\sim\!3{,}500\,\K$ electron temperature of the GCL against the values of $T_e$ for the Galactic \hii\ regions from \citet{balser16} in Figure~\ref{fig:te_scatter}.  This figure shows that the electron temperature of the GCL, indicated by the horizontal red line, is anomalously low compared to that of Galactic \hii\ regions, for all Galactocentric radii.  The computation of $T_e$ assumes LTE, but departures from LTE would result in overestimates of the true value of $T_e$ \citep[e.g.][]{balser24}.

In Figure~\ref{fig:te_comparison}, we compare the values in the electron temperature map from Paper~I and the Hydrogen RRL FHWM-derived maximum $T_e$ values from the single Gaussian fits in order to determine if the low electron temperature and line widths agree with each other.  This figure shows that although the values of $T_e$ are low, they are consistent with the maximum values derived from the RRL FWHM analysis.
The magnitude of the turbulent broadening in the GCL is generally only $\sim\!5$ to 10\,\kms when the linewidth is computed for individual spaxels.  If instead we integrated over the entire source, the measured hydrogen RRL line width of 20.5\,\kms (see Section~\ref{sec:linewidth}) would imply turbulent broadening of 16.1\,\kms, assuming $T_e = 3500\,\K$.



\subsection{Association with stellar clusters and OB stars}

\begin{figure}
    \centering
    \includegraphics[width=3.25in]{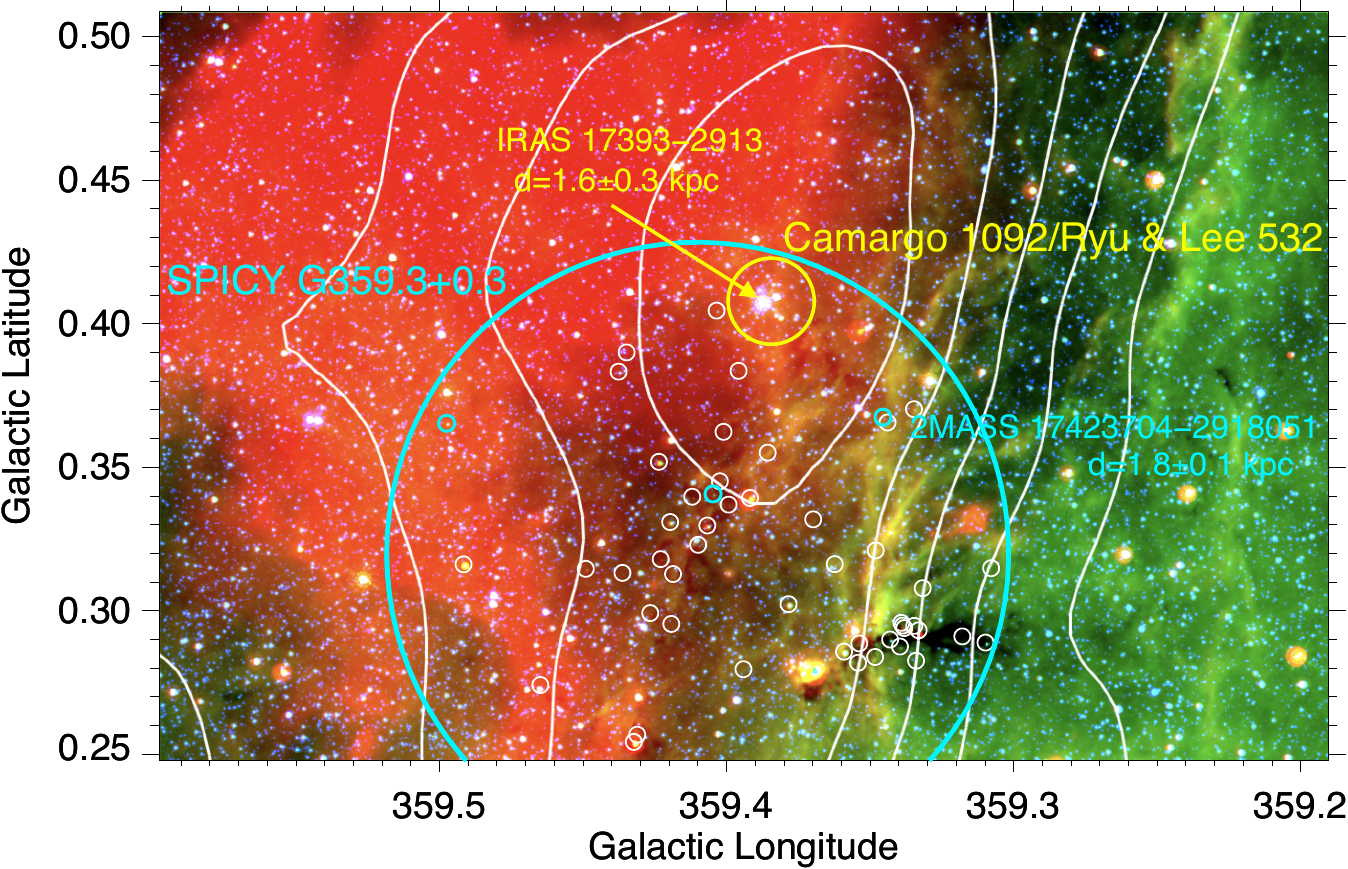}
    \caption{Two recently identified cluster candidates toward the GCL: stellar cluster candidate \citet{camargo16} \#1092/\citet{ryu18} \#532 and young stellar object (YSO) cluster candidate SPICY G359.3+0.3 \citep{kuhn20}.  The background image shows Spitzer MIPSGAL 24.0\,\micron\ data in red, Spitzer GLIMPSE 8.0\,\micron\ data in green and Spitzer GLIMPSE 3.6\,\micron\ data in blue, with contours of integrated GDIGS RRL emission as in Figure~\ref{fig:gc}.  The stellar cluster candidate is located at the peak of RRL emission and contains the red supergiant candidate IRAS 17393$-$2913 that has a GAIA eDR3 parallax distance $d=1.56^{+0.26}_{-0.20}\,\kpc$. The YSO cluster contains a Class II YSO candidate at a distance $d =1.82^{+0.10}_{-0.09}$\,\kpc.\label{fig:stellar_counterparts}}
\end{figure}

If the GCL is an \hii\ region, we may be able to identify the  ionizing source(s) and/or active star formation within it.  We search the GCL area for evidence of stellar clusters and young stellar objects, and show the location of identified objects in Figure~\ref{fig:stellar_counterparts}.

Two independent investigations have identified a cluster candidate within the GCL: \citet{ryu18} cluster candidate \#532 and \citet{camargo16} cluster candidate \#1092. Both of these cluster candidates are located at $(\ell,b)=(-0.61\degree,0.41\degree)$, near the RRL intensity peak (see Figure~\ref{fig:stellar_counterparts}).
Within the cluster candidate the brightest infrared source is IRAS 17393$-$2913 (also designated as 2MASS ~ J17423377$-$2914392 and MSX6C~G359.3869+00.4081). The source is fainter at shorter relative to longer wavelengths, as it has (Vega) magnitudes ($V$, $G$, $J$, $H$, $K_{s}$, ${\rm IRAC~4.5\,\micron},$ ${\rm MSX~8.3\,\micron}) = ($15.90, 11.63, 5.49, 3.91, 3.30, 3.46, 2.34).  Gaia eDR3 lists a parallax for IRAS 17393$-$2913 of $\varpi=0.6135 \pm 0.0922$\,mas, with astrometric excess noise of 0.7\,mas and renormalized unit weight error (RUWE) of 1.0. Using the code of \citet{lindegren20} to calculate the zero-point offset (and overriding the warning that ${\tt nu\_eff\_used\_in\_astrometry}$ is slightly out of range) yields a zero-point offset of $-0.027$\,mas and a corrected parallax of $\varpi=0.641 \pm 0.0922$\,mas, or a distance of $d=1.56^{+0.26}_{-0.20}\,\kpc$. 
With no extinction correction, 
the absolute $K$ magnitude of IRAS 17393$-$2913 is then $M_{K}=-7.67$.

There are no catalogued OB stars in the vicinity of the RRL intensity peak and only four contained within the total area of the GCL. Using the predicted absolute magnitudes of O-stars in \citet{martins06} an O3V star would have an apparent $K$-band magnitude of $K=5.9$ to 6.5 and an O9V star would have $K=7.6$ to 8.2 at distances from 1.5 to 2\,kpc. Allowing for moderate extinction, 
we identify three candidate O-stars near the RRL intensity peak whose GAIA eDR3 parallaxes and photometry make them candidates for OB stars  $\sim\!2\,\kpc$ from the Sun: 2MASS J17424931$-$2912223 ($\varpi=0.4729\pm
0.02664$; $K_s = 8.07$), 2MASS 17422194$-$2911134 ($\varpi = 0.6609\pm
0.1153$; $K_s = 8.05$), and 2MASS 17425287$-$291829 ($\varpi = 0.4591\pm
0.1090$; $K_s = 8.09$).  


Separated from the above cluster by 6.5\arcmin\ is a recently identified cluster of young stellar objects (YSOs) in the Spitzer IRAC Candidate YSO (SPICY) catalog \citep[][see Figure~\ref{fig:stellar_counterparts}]{kuhn20}.  This cluster, SPICY~G359.3+0.3, consists of 50 sources within a radius of $5.4\arcmin$, three of which have Gaia parallaxes. The distance estimate, principally determined by the class II YSO candidate 2MASS J17423704$-$2918051 with zero-point corrected Gaia eDR3 parallax $\varpi=0.549 \pm 0.028$\,mas, is $d=1.82_{-0.09}^{+0.10}$\,\kpc. The two other cluster members have zero-point corrected Gaia parallaxes that correspond to distances of $1.3_{-0.5}^{+2.7}$\,\kpc\
(2MASS 17425187$-$2915544) and $d>1.9$\,\kpc\ (2MASS 17425953$-$2910235).  These distances are marginally consistent with the distance to IRAS 17393$-$2913 given above.


\section{Discussion and Summary}

If the GCL is foreground to the Galactic center, as argued in  Paper~I, we hypothesize that it is an \hii\ region.  We found here that the GCL has a similar MIR morphology, MIR flux ratio, size (for reasonable distance estimates), and RRL FWHM to that of the known population of Galactic \hii\ regions.  Furthermore, we identify here the possible ionizing cluster at the RRL peak and also find a nearby cluster of YSOs.  Cluster members with Gaia parallaxes have distances of $1.7\pm 0.4$\,\kpc.  If the cluster of YSOs is associated with the GCL, that would indicate ongoing star formation.


The low electron temperature values of the GCL remain a confusing aspect of its classification as an \hii\ region.  If the GCL were an \hii\ region located a couple kpc from the Sun, we would expect it to have an electron temperature of approximately 8{,}000\,\K instead of the $\sim\!3{,}500\,\K$ value that we find.  

Despite having a value of $T_e$ that is unusual for a Galactic \hii\ region, the weight of the evidence is consistent with the GCL being a foreground \hii\ region.
In our experience, the MIR and radio morphology of the GCL is characteristic of evolved \hii\ regions that have large physical sizes.  If it is 2\,\kpc\ away, the GCL would have a physical diameter of $\sim35\!\pc$, which is on the larger end of the Galactic \hii\ region size distribution.  
\begin{acknowledgments}
We thank the referee for insightful comments that helped the clarity of this manuscript. 
\nraoblurb\  We thank West Virginia University for its financial support of GBT operations, which enabled some of the observations for
this project.  This work is supported by NSF grant AST1516021 to LDA,  NASA grant NNX17AJ27GRAB to RAB, and Australian Research Council (ARC)  grant DP160100723 to NM-G. NHW is supported by an Australian Research Council Future Fellowship (project number FT190100231) funded by the Australian Government.  Some of this work took part under the program SoStar of the PSI2 project funded by the IDEX Paris-Saclay, ANR-11-IDEX-0003-02. This research has made use of the SIMBAD database, operated at CDS, Strasbourg, France.  
\end{acknowledgments}

\bibliographystyle{apj}
\bibliography{ref.bib}

\end{document}